\documentclass[useAMS,usenatbib]{mn2e}
\usepackage{graphics,epsfig,psfig}
\usepackage[normalem]{ulem}
\usepackage[dvipsnames]{color}
\usepackage[]{inputenc,amssymb}

\def \be{\begin{equation}}
\def \ee{\end{equation}}
\def \bea{\begin{eqnarray}}
\def \eea{\end{eqnarray}}
\def\etal{{et al.\ }}

\title[Superbubble breakout and galactic winds from disk galaxies]{
Superbubble breakout and galactic winds from disk galaxies
}

\author[Arpita Roy, Biman B. Nath, Prateek Sharma, Yuri Shchekinov]
{Arpita Roy$^{1,2}$ \thanks{arpita@rri.res.in}, Biman B. Nath$^1$, Prateek Sharma$^2$,
Yuri Shchekinov$^3$\\
$^1$Raman Research Institute, Sadashiva Nagar, Bangalore 560080, India\\
$^2$Joint Astronomy Programme and Department of Physics, Indian Institute of Science, Bangalore 560012, India\\
$^3$ Department of Physics, Southern Federal University, Rostov on Don 344090, Russia\\
}

\begin{document}


\maketitle

\label{firstpage}

\begin{abstract}
We study the conditions for disk galaxies to produce superbubbles that can break
out of the disk and produce a galactic  wind. We argue that the threshold surface density of supernovae rate for seeding a wind depends on the ability of  superbubble  energetics to compensate for radiative cooling.
We first adapt Kompaneets formalism for expanding bubbles in a stratified medium to the case of continuous energy injection and include the effects of radiative cooling in the shell.
With the help of hydrodynamic simulations, 
we then study the evolution of superbubbles evolving in stratified disks with typical disk parameters. We identify two  crucial energy injection  rates that differ in their effects, the corresponding breakout ranging from being gentle to a vigorous one. (a)  Superbubbles that break out of the disk with a Mach number  of order $2\hbox{--}3$ correspond to an energy injection rate of order $10^{-4}$ erg cm$^{-2}$ s$^{-1}$, which is relevant for disk galaxies with synchrotron emitting gas in the extra-planar regions. (b) A larger energy injection threshold, of order $10^{-3}$ erg cm$^{-2}$ s$^{-1}$, or equivalently, a star formation surface density of $\sim 0.1$ M$_{\odot}$ yr$^{-1}$ kpc$^{-2}$, corresponds to superbubbles with a Mach number  $\sim 5\hbox{--}10$. 
While the milder superbubbles can be produced by large OB associations, the latter kind requires super-starclusters.
These derived conditions compare well with observations of disk galaxies with winds and the existence of multiphase halo gas. Furthermore, we find that contrary to the general belief that superbubbles 
fragment through Rayleigh-Taylor (RT) instability when they reach a vertical height of order the scale height, the superbubbles are first affected by thermal instability for typical disk parameters and that RT instability takes over when the shells reach a distance of approximately  twice the scale height. \end{abstract}


\section{Introduction}
Observations of nearby and high-redshift galaxies have shown that star formation in them often leads to galactic winds. Starburst galaxies, with star formation rate (SFR) in excess of
a few tens of $M_{\odot}$ yr$^{-1}$ are known to excite such outflows. However, Heckman (2002) pointed out that it is not the average SFR, but the SFR surface density  which  is a deciding factor for
the existence of outflows. He found a threshold SFR surface density of $\sim 0.1$ M$_{\odot}$
kpc$^{-2}$ yr$^{-1}$ as a prerequisite for starbursts to be able to produce galactic winds.

The standard scenario of star formation leading to the wind phenomena posits that super-starclusters give rise to a large number of supenovae (SN) in a relative small region,  which can  produce a 
superbubble in the disk and can break out of the disk with enough momentum to produce a wind. Such super-star clusters, or young globular clusters, have been observed to have masses in the range of few
$\times 10^5\hbox{--}6 \times 10^7$ M$_{\odot}$ within a typical radius of $\sim 3\hbox{--}10$ pc (Ho 1997; Mart\'in-Hern\'andez \etal 2005, Walcher \etal 2005). The large amount of energy deposited into the 
interstellar medium (ISM) by these objects in the form of UV radiation and mechanical energy is believed to be an important feedback process. The mechanical energy from these super-starclusters has been shown to 
be important for the superbubble produced by the combined SNe to break out of the disk and produce a large scale wind (e.g.,
Tenorio-Tagle, Silich, Mu\~{n}oz-Tu\~{n}\'on 2003). 

There have been a number of calculations, both analytical and numerical, dealing with the breakout of superbubbles from disk galaxies. The conditions for breakout depend strongly on the assumption of the stratification 
of gas in the disk. Consider an exponentially stratified disk with mid-plane ambient gas pressure $P_0$, gas density $\rho_0$, scale height $z_0$, and a bubble being blown by mechanical luminosity $\mathcal{L}$.
Mac Low \& McCray (1988) defined a dimensionless parameter  $D\equiv {\mathcal{L} \rho_0^{1/2} /( P_0^{3/2} z_0^2})$ , and noticed in their numerical simulations that breakout of bubbles occurred when $D \ge 100$.  The 
importance of this parameter can be understood by considering the self-similar evolution of a superbubble driven by an energy injection rate of $\mathcal{L}$, given by $r \sim ({\mathcal L} t^3/\rho_0)^{1/5}$, and
$\dot{r} \sim (3/5) ({\mathcal L} /\rho_0)^{1/5} t^{-2/5}$. This implies a speed of $\sim (3/5) ({\mathcal L} /\rho_0 z_0^2)^{1/3} \propto D^{1/3}$ when the superbubble reaches a distance of the scale height, for an 
ambient gas at a given temperature.
According to this criterion, for a scale height $z_0=200$ pc, and mid-plane gas  density $\mu_H n_0\sim2.3 \times 10^{-24}$ g cm$^{-3}$, $P_0/k_b \sim n_0 10^4$ K cm$^{-3}$, a bubble with total mechanical luminosity 
of $\mathcal{L} \sim 3.8 \times10^{37}$ erg~s$^{-1}$ will be able to breakout of the ISM.

Basu \etal (1999) defined a dimensionless parameter $b\equiv (27/154 \pi)^{1/2} \mathcal{L}^{1/2} \rho_0^{1/4} P_0 ^{-3/4} z_0 ^{-1}$ which is a ratio 
of the radius where the Mach number of the superbubble becomes unity, to the scale height. This is motivated by the self-similar solution of a stellar 
wind, $r \sim (125/154 \pi)^{1/5}\mathcal{L}^{1/5}\rho_0^{-1/5} t^{3/5}$. They showed that this parameter is related to the above mentioned $D$ parameter as $D=17.9 b^2$. In other words, a superbubble with $b < 1$ is 
likely to be confined where as blowout will occur for $b \ge 1$.

Koo \& McKee (1992) analytically determined a  condition for the breakout. Since the bubbles accelerate after reaching a distance of order the scale height, owing to the rapidly decreasing density, it becomes liable to fragment due to Rayleigh-Taylor instability. If the Mach number of the bubble  at scale height is $\ge 3$, then they argued that the bubble would be able to breakout. 
They used radiative  bubble model of Weaver et al (1977) for a uniform density atmosphere in order to derive a critical mechanical luminosity for which the Mach number is unity, $\mathcal{L}_{cr}\sim  17.9 \rho_0 z_0^2 c_s^3$, where $c_s$ is the isothermal sound speed of the ambient gas. The Mac Low \& McCray condition of $D \ge 100$ translates to $\mathcal{L}/\mathcal{L}_{cr} \ge 5$.  As we will find in our simulations, the Mach number of a bubble after breakout is of order  $(1/ 5 c_s)( {\mathcal L  / \rho_0 z_0^2} )^{1/3}$. Therefore, the Mac Low-McCray condition of $D \ge 100$
translates to the condition that the Mach number at breakout is of order unity. We also note that they considered superbubbles that originated at a height from the mid-plane, which made it easier for bubbles to break out. Our simulations show that the critical luminosity for Mach number at a distance of the scale height to be unity is ${\mathcal L}_{cr}\sim 125 \rho_0 z_0^2 c_s^3$, larger
than the estimate of Koo \& McKee (1992).

Koo \& McKee (1992) then considered an additional strata of HII gas with a scale height of $1$ kpc and mid-plane number density $0.025$ cm$^{-3}$, and found the breakout condition to be of order $N_{OB}\sim 800$, or equivalently, $\mathcal{L} \ge 4.1 \times 10^{38}$ erg~s$^{-1}$. Silich \& Tenorio-Tagle (2001) considered the effect of halo gas pressure and determined a minimum energy for the superbubble  to blow out of the galaxies (with both disk and spherical ISM distribution)
 with ISM gas mass in the range of $10^6\hbox{--}10^9$ M$_{\odot}$. For a disk galaxy with $M_{ISM}\sim 10^9$ M$_{\odot}$, they found a minimum energy of $\sim 10^{38}$ erg~s$^{-1}$, corresponding to $N_{OB} \sim 1000$.

 As Heckman (2002) has emphasized, it is the surface density of SFR that determines the condition for the existence of galactic winds, and not the total luminosity. To translate the above energy conditions into a surface density, we need to estimate the surface area of such bubbles at the breakout epoch. In this paper, we re-visit this issue in order to understand the empirical threshold SFR surface density for galactic winds. Murray, M\'enard, Thompson (2011) have recently argued that radiation pressure from UV radiation from a disk with a SFR surface density larger than $0.1$ M$_{\odot}$ kpc$^{-2}$ yr$^{-1}$ can produce a large scale wind. This estimate however crucially depends on the assumption of the grain opacity, and as Sharma \& Nath (2012) have shown the relevant opacity at UV may fall short of the requirements.

There have also been studies on the existence of multiphase gas in the halos of spiral galaxies,
and their connection to the star formation properties in the disk.  Dahlem, Lisenfeld, Golla (1995) 
considered nine edge-on galaxies with extended synchrotron emitting halo gas, and derived a minimum value of surface density of energy injection for superbubble breakout, as $\sim 10^{-4}$ erg~s$^{-1}$ cm$^{-2}$. T\"ullmann \etal (2006) further considered X-ray, radio and far-infrared (FIR) emission from the extended halo gas in a sample of 23 edge-on spiral galaxies, and found that the halo contained gas at low and high temperatures (multiphase)
if the surface density of energy injection in the disk exceeds $\sim 10^{-3}$ erg~s$^{-1}$ cm$^{-2}$.
If the existence of multiphase halo gas depends on the process of superbubbles breaking out of the disk and depositing hot interior gas ( as suggested by Tomisaka \& Ikeuchi 1986; Tenorio-Tagle, Rozyczka, Bodenheimer 1990), as well as cold gas in the shell, then it would be interesting to compare the energetics of such superbubbles and the observed threshold energy injection rate.

In this paper, we study the standard scenario of thermal pressure of the gas interior to superbubbles being the driving mechanism for the wind, and derive a threshold condition for the superwind. We find that radiative loss of energy is important for the dynamics of shocks, and the inclusion of radiation loss increases the energy budget for the bubbles to breakout of the disk and
produce a wind. We also find that our estimate of the threshold energy requirement can explain the observed threshold SFR surface density for galactic outflows.

The paper is organized as follows. In \S 2 we derive an order-of-magnitude estimate of the threshold based on the key idea that the superbubble energetics needs to balance radiative cooling. Then we present the analytical formalism in \S 3 and discuss the results in \S 4. We then present the results from numerical simulations in \S 5, and discuss the effect of thermal and RT instability in \S 6.

\section{Analytic Estimates}
To begin with, we derive a threshold rate of SNe for a superbubble to continue to grow and ultimately breakout of the disk from simple arguments. We can first consider the condition that
the superbubble is able to drive a strong shock in the disk. This requires the volume energy injection time scale  to be shorter than the sound crossing time. In other words, if we consider a region of radius $R$ in the disk and an energy  injection rate of $\mathcal{L}$, then one needs
\be
 { 1.5 n kT \over\mathcal{L}/(4 \pi R^3/3) } \ll R/c_s \,,
 \ee
  where $c_s$ is the sound speed. 
This gives a lower limit of $\mathcal{L}/(\pi R^2) \gg 3 \times 10^{-6} n T_4^{3/2} $ erg~s$^{-1}$ cm$^{-2}$, where $n$ is the ambient gas particle density in cm$^{-3}$, and $T=T_4 \, 10^4$ K.

 A second, and more stringent, constraint on SNe luminosity comes from accounting for radiative losses.
 Let us assume that  when a SN remnant enters the radiative stage it quickly
loses its energy and does not contribute to the energy input  of the superbubble. Assume then that the radiative stage begins when the post-shock  temperature is 
$T_s \simeq2 \times 10^5$ K such that the radiation loss function  is maximum and much larger than 
the minimum at $\sim 10^6$ K. We therefore define the time when a SN remnant loses its 
energy at time when the shock velocity is $v_s=120$ km s$^{-1}$ (corresponding to the post-shock temperature of $2 \times 10^5$ K). It determines the corresponding time and radius as 
(see also Kahn 1998, who defined this as the beginning of phase III in the evolution of a bubble),
\be
t_a=1.4 \times 10^5{E_{51}^{1/3}\over n^{1/3}}~{\rm yr}, ~~~R_a=37{E_{51}^{1/3}\over n^{1/3}}~{\rm pc}\,.
\ee

One can therefore define the coherency condition as,
\be
{4\pi\over 3}R_a^3t_a\nu_{_{SN}}> 1 \,,
\ee
which means that before a SN remnant stalls because of cooling losses,  another SN explosion injects energy into the remnant 
and forms a single bubble. This condition determines the required SN rate 
\be
\nu_{_{SN}}>30\times 10^{-12}\left({n\over E_{51}}\right)^{4/3}~{\rm SNe}~{\rm yr}^{-1}~{\rm pc}^{-3} \,.
\ee
We can estimate the surface density of SNe, by multiplying this rate density by the scale
height, which is the height of a bubble at the epoch of breakout. For a scale height of  $500 \, z_0=z_{0,0.5}$ pc,  this corresponds to   $1.5 \times 10^{-2} (n/E_{51})^{4/3} \, z_{0,0.5} \, {\rm SNe~{\rm yr}^{-1}}~{\rm kpc}^{-2}$. (The scale height is relevant here because, as we will see later, the maximum radius of bubbles in the plane parallel to the disk is of order $\pi z_0$.) Finally, we recall that for a Salpeter IMF, one SN corresponds
to $150$ M$_{\odot}$ of stellar mass, considering stars in the range of $1\hbox{--}100$ M$_{\odot}$. Therefore the threshold condition for SFR surface density becomes       $\sim 2.5 (n/E_{51})^{4/3}  z_{0,0.5} {\rm M}_{\odot}~{\rm yr}^{-1}~{\rm kpc}^{-2} $. The corresponding surface density of energy injection is $\sim 0.05 \, n^{4/3} \, E_{51}^{-1/3}  \, z_{0,0.5} $ erg~s$^{-1}$ cm$^{-2}$.
It is interesting to find that these above estimates of the threshold energy injection or SFR surface density
 are comparable to the observed threshold for the
existence of multiphase halo gas (T\"ullmann \etal 2006) and  superwinds  (Heckman 2002).

\section{Kompaneets approximation}
We first discuss the expansion of blastwaves in a stratified atmosphere, in the adiabatic case and then for radiative shocks. Kompaneets (1960) had first analytically worked out the case of adiabatic shocks  in this case (see, e.g., Bisnovatyi-Kogan \& Silich 1995). Consider an exponentially stratified medium described by ${\rho{(z)}}={\rho_{0}\exp{(-z/z_{0})}}$, where $\rho_{0}$ is the midplane density and $z_{0}$ is the scale height and $E_{0}$ is the explosion energy. It is assumed that the shock pressure is uniform, and is given by,
\begin{equation}
 P_{sh} ={(\gamma-1) \lambda E_{0}\over V}\,,
\label{eq:shilpi2}
\end{equation}
where $\lambda\sim 1$ (Kompaneets 1960) is a constant that differentiates the shock pressure from the
average pressure inside the bubble; Bisnovatyi-Kogan \& Silich (1995) evaluated $\lambda=1.33$.
We use $\lambda=1$ for simplicity.
We define a dimensionless time-like parameter as,
\begin{equation}
{y}={\int_{0}^{t}\sqrt{{{(\gamma^2-1)}  E_{0}} \over {2\rho_{0}V}}dt}\,.
\label{eq:komp6}
\end{equation}
Where $E_{th}$ is the thermal energy of the interior gas, $V$ is the volume of the bubble and $t$
is the time. 
 The shape of the shock front is derived as,
\begin{equation}
 r=2z_0\arccos \Bigl \{{{1 \over 2}\exp{(z/2z_0)}{\Bigl [1-{y^2 \over 4z_0^2} +\exp{(-z/z_0)} \Bigr ]}} \Bigr \}\,.
\label{eq:komp14}
\end{equation}
The location of the top and bottom of the bubble then follows by setting $r=0$ ( with $\tilde{y}=y/z_0$),
\begin{equation}
{z_\pm {(\tilde{y})}} = {-2z_0.\ln{(1\mp \tilde{y}/2)}}\,,
\end{equation}
which shows that the top of the bubble reaches infinity when $y \rightarrow 2z_0$ while $t$
remains finite. This implies that the bubble accelerates in the $z$-direction due to stratification,
after an initial deceleration phase when the bubble is small and spherical, as in the usual Sedov-Taylor solution. The maximum cylindrical radius of the bubble is also obtained from the above solution by putting $(\partial r /\partial z) =0$,
\begin{equation}
{r_{\rm max}{(\tilde{y})}}={2z_0\arcsin {(\tilde{y} /2)}}\,.
\label{eq:rmax}
\end{equation}
The $z$-component of the velocity of the topmost point of the bubble is given by,
\begin{equation}
{v_{z}{(\tilde{y})}}={{1 \over {(1-\tilde{y}/2)}}\sqrt{{{(\gamma^2-1)} \over 2}{ E_{0} \over \rho_{0}V{(t)}}}}\,.
\label{eq:komp18}
\end{equation}

\subsection{Continuous energy injection}
We  can extend Kompaneets approximation and radiative blastwave calculation to the case of
continuous energy injection. Schiano (1985) had done a similar calculation in the case of an active galactic nucleus. Consider an association with $N_{OB}$ stars with masses above
$8$ M$_{\odot}$, which ultimately produce supernovae. 
If we consider the main-sequence lifetime as $\tau_{SN} \sim 5\times10^7$ yr for these stars, then
the total mechanical luminosity of the SN in the association can written as, 
\begin{equation}
 \mathcal{L}= 6.3 \times 10^{35} N_{OB} \, E_{51} \, (\tau_{SN}/5 \times 10^7 {\rm yr}) ^{-1}
  \quad{\rm  erg} \, {\rm s}^{-1}\,,
\label{eq:kompaneets6}
\end{equation}
where supernova energy is $10^{51} E_{51}$ erg.  As McCray and Kafatos (1987) have argued, since the main sequence life time scales with the stellar mass as $\tau \propto M_\ast ^{-1.6}$,
and since the initial mass function (IMF) is given by, $dN_\ast/ d(\log M_\ast) \propto M_\ast^{-1.35}$, for a Salpeter  IMF, the rate of SN will scale with time as $\propto
{dN_\ast \over d M_\ast} {dM_\ast \over dt}\propto t^{2.35/1.6} \, t^{-1/1.6-1}\propto t^{1.35/1.6-1}$, which is
roughly constant in time. Here we have used ${dM_\ast \over dt} \propto t^{-1/1.6-1}$, given the above
mentioned dependence of stellar main sequence lifetime.
Therefore we can write, for the adiabatic case, the total energy in the
superbubble as $E_{th} = \mathcal{L} t$.

\begin{figure}
\centerline{
\epsfxsize=0.5\textwidth
\epsfbox{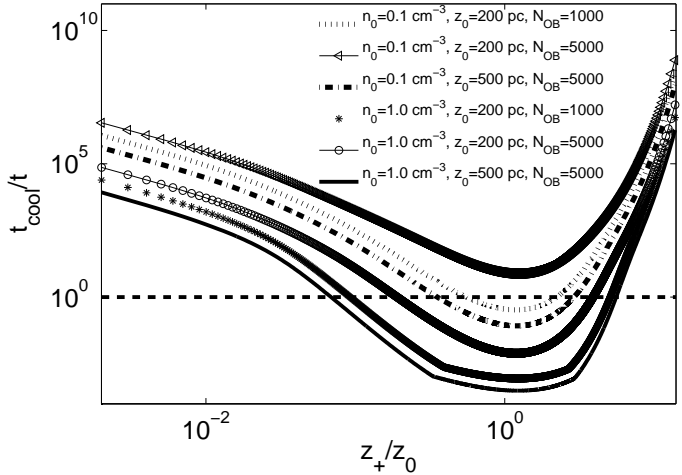}
}
\caption{
The ratio of cooling time to time ($t_{\rm cool}/t$) is plotted against the height of adiabatic superbubble with continuous energy injection, for different combinations of $N_{OB}, n_0$, and $z_0$.}
\label{fig:tcool}
\end{figure}

Instead of eqn \ref{eq:komp18}, the $z-$velocity of the
top of the bubble is then given by,
\begin{equation}
v_z (\tilde{y})
={1 \over (1-\tilde{y}/2)} \sqrt{ {(\gamma^2-1) \over 2}{\mathcal{L}t  \over \rho_0 V(t)} }
\,,
 \label{eq:komp31}
\end{equation}
and the corresponding $y$ parameter is also written in terms of $t$, as
\begin{equation}
y
=\int_0^t \sqrt{ {(\gamma^2-1)  \mathcal{L}t' \over 2 \rho_0 V(t')}} \, dt' 
\,.
 \label{eq:komp32}
\end{equation}

These equations can determine the dynamics of the superbubble in the case of continuous energy injection.

\subsection{Radiative loss with continuous injection}
Radiative losses can be important for the dynamics of  both the blastwave and a superbubble with continuous energy injection.  Shocks become radiative when
the cooling time $t_{\rm cool} \ll t$. The cooling time behind the shell can be estimated as $t_{\rm cool}=1.5kT /( 4n\Lambda{(T)})$, for a strong shock with $n=n_0 \exp(-z/z_0)$, and the shock temperature being estimated from the
shock speed (in the $z-$direction, say). We assume a cooling function, as given by Eqn 12 in Sharma \etal 2010, appropriate for 
gas with $10^4
\le T(\equiv 10^6 T_6 )  \le 10^7$ K and given as follows:
\begin{eqnarray}
\Lambda (T)&=10^{-22}(8.6 \times 10^{-3} T_{keV}^{-1.7} + 5.8 \times 10^{-2} T_{keV}^{0.5} +\nonumber\\ 
& 6.3 \times 10^{-2}) \,\, erg\,s^{-1} cm^3 \, , \, \, T>0.02 \, keV.\nonumber\\
&=6.72 \times 10^{-22}(T_{keV}/0.02)^{0.6} \,\, erg\,s^{-1} cm^3 \, ,\, \nonumber\\
&  \, T \leq 0.02 \, keV, \, T \geq 0.0017235\, keV.\nonumber\\
&=1.544 \times 10^{-22}(T_{keV}/0.0017235)^{6.0} \,\, erg\,s^{-1} cm^3\, ,\, \nonumber\\
&  \, T < 0.0017235 \, keV \, ,
\label{eq:cooling_fn_formula}
\end{eqnarray}
where T$_{keV}$ is the temperature in keV.
Figure \ref{fig:tcool} shows  the ratio $t_{\rm cool}/t$ as a function of the bubble height $z_+$
for 
bubbles with continuous energy injection for a few cases. 
The curves show that 
the shock enters the radiative phase much before reaching the scale height 
unless the ambient density and scale height are very small and $N_{OB}$ is very large (e.g., the case with
$n_0=0.1$ cm$^{-3}$, $z_0=200$ pc, $N_{OB}=5000$).

Radiation loss from the shocked medium can therefore be important (see also Maciejewski \& Cox 1999). Kovalenko \& Shchekinov (1985) had calculated the dynamics of a blastwave with radiative loss, assuming that  the shock kinetic energy  is converted into thermal energy of gas in a thin shell behind it, and that radiative loss from this shell keeps
the shock isothermal. It can then be shown that for a strong shock the energy lost per unit mass is $\sim (1/2) u_s^2$, where $u_s$ is the shock speed. From the Hugoniot condition for a strong shock that,
\begin{equation}
 {u_s^2}={{(\gamma+1)}P_{s} \over 2\rho_{0}} ={{(\gamma^2-1)} E_{th} \over 2\rho_{0}V{(t)}} \, ,
\label{eq:komp20}
\end{equation}
where $E_{th}$ is the thermal energy of the shocked gas. The structure of the shock in this case is such that the interior gas remains hot and adiabatic, whereas the shocked ambient gas that is swept into a shell loses its energy radiatively and is kept at a constant temperature ( at $\sim 10^4$ K). We note that Mac Low \& McCray (1988) showed that the radiative loss from the interior hot gas of the bubble does not change the dynamics of the bubble.

Following the calculation of Kovalenko \& Shchekinov (1985) for a radiative blastwave, we assume that bubbles with continuous
energy injection also form an isothermal thin shell,
after a certain time $t_1$ when it enters the radiative phase. For simplicity, we also assume a self-similar solution for a spherical shock, of the type given by Weaver \etal (1977), $r_s = A \mathcal{L}^{1/5} t^{3/5}$, where $A$ is a constant depending on the ambient density.
Furthermore, Weaver \etal (1977) have pointed out that a fraction $6/11$ of the total energy is stored
in the shell and the rest in the rarefied gas inside. In the spirit of Kovalenko \& Shchekinov (1985) we assume the total shell energy to be thermal in nature. In other words,
initially $E_{th}=(6/11) \mathcal{L} t$. We can determine the time evolution of $E_{th}$ as follows.

 Using the result derived in eqn \ref{eq:komp20} that the amount of energy lost  per unit volume is 
$(1/2) \rho_0 u_s^2= (\gamma ^2 -1)  E_{th}/(4V (t))$,
 we can  write for the evolution of thermal
energy in this case, 
\begin{eqnarray}
{E_{th}{(r)}} =& &{6 \over 11} {\mathcal{L}t-\pi{(\gamma^2-1)}\int_{r_{1}}^{r}{ E_{th}(r) \over V}r^2dr}\,,\nonumber\\
=&& {6 \over 11}  {\mathcal{L}^{2/3}\,\Bigl ({r \over A}\Bigr )^{5 /3} -\pi{(\gamma^2-1)}\int_{r_{1}}^{r}{ E_{th}(r) \over V}r^2dr}\,.
 \label{eq:komp33}
\end{eqnarray}
\begin{figure}
\centerline{
\epsfxsize=0.5\textwidth
\epsfbox{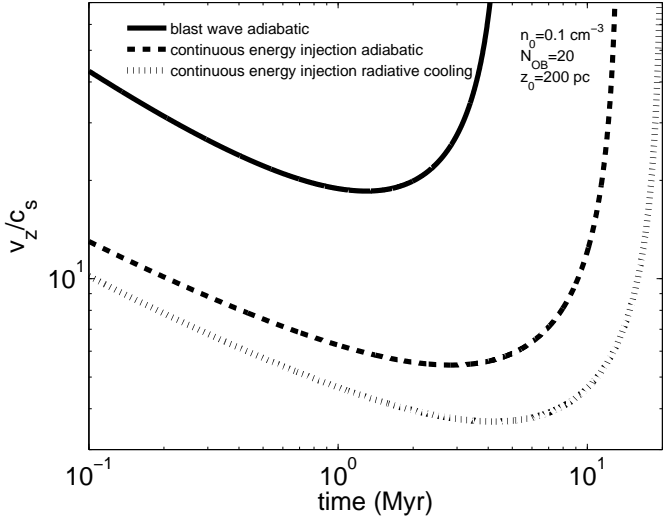}
}
{\vskip-3mm}
\caption{
The evolution of the ratio of $v_z$ to $c_s$ (the sound speed for an ambient gas at $10^4$ K) is plotted against time, for an adiabatic blastwave (thick solid line), adiabatic superbubble with continuous energy injection (dashed) and with radiative loss (dotted line).
}
\end{figure}
Here  $r_1$ is the radius at time $t_1$.
 We can explicitly solve this equation for a spherical shock, and then use the results to estimate
the $z-$ velocity of an oval shaped bubble. For a spherical shock (with volume $ V={4\over 3} \pi r^3$), the energy equation (no. \ref{eq:komp33})  can be shown to yield a solution of the type $E_{th}{(r)}= br^\alpha $,
where
\begin{equation}
 b[\alpha+{3\over4}(\gamma^2-1)]r^{\alpha -1} = {6 \over 11} {5 \over3}{\mathcal{L}^{2/3} \over A^{5/3} }\, r^{2/3}  
 \,.
\label{eq:kompaneets11}
\end{equation}
Comparing the powers of $r$ from both the sides we get,
$ \alpha = {5 \over 3} $. Putting this value of $\alpha$ in eqn \ref{eq:kompaneets11} and comparing the coefficients of time on both  sides we get, 
\begin{equation}
 b={30 \over 99}{\mathcal{L}^{2/3} \over A^{5 /3}}
 \,.
\label{eq:kompaneets12}
\end{equation}
Therefore $E_{th}(r)$ becomes,
\begin{equation}
 E_{th}(r)={0.3 \mathcal{L}} t 
 \,,
\label{eq:kompaneets13}
\end{equation}
 showing that roughly 70\% of the total energy is radiated away.
 Note that this  is an asymptotic value of the loss in the limit $r \gg r_1$, in the regime where the approximation $E \propto r^\alpha$ is valid. 
 We can therefore use equations \ref{eq:komp31} and \ref{eq:komp32}, with the above
 value of
$E_{th}$, and determine the dynamics of a radiative superbubble with continuous energy injection.

\section{Analytic results}
Figure 2 shows the evolution of the Mach number for a $10^4$ K gas as a function of time,  for an adiabatic blastwave, a superbubble with continuous energy injection with and without radiative loss. It is convenient to 
define a dynamical time scale for this problem (Mac Low \&
McCray 1988), $t_{d}\sim z_0^{5/3} (\rho_0/\mathcal{L})^{1/3}$, which is the expected time to reach the scale height for a self-similar evolution of superbubbles.
For $z_0=200$pc, $\mathcal{L}\sim 1.3 \times 10^{37}$ erg~s$^{-1}$ and $\rho_o\sim 10^{-25}$ g cm$^{-3}$
(for $\mu \sim 0.6$) ,  $t_d \sim 2.8$ Myr. 
We find that the $z-$velocity shows a minimum at $\sim 1.5 t_d$, when it reaches a distance of the
scale height.  We denote this  minimum value of $z-$velocity as $v_{z,{\rm min}}$, and refer to this epoch as the 'stalling epoch' in our discussion below.

\begin{figure}
\centerline{
\includegraphics[scale=0.35,angle=-90]{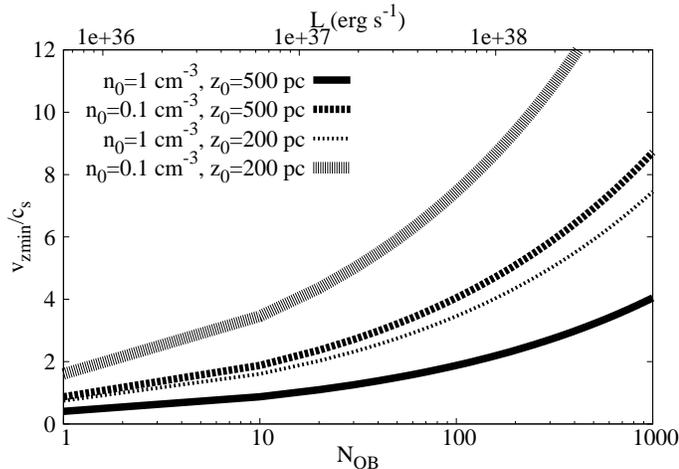}
}
{\vskip-3mm}
\caption{
The ratio $v_{z,{\rm min}}/c_s$ of the $z-$velocity of the top of the bubble to the sound speed of the ambient gas at $10^4$ K is plotted as a function of $\mathcal{L}$ the mechanical luminosity, and $N_{OB}$, the number of SNe responsible for the bubble. Different lines correspond to
different values of mid-plane gas} number density $(1,0.1)$ cm$^{-3}$ and scale heights ($200,500$) pc.
\label{fig:komp_res}
\end{figure}

Figure \ref{fig:komp_res} shows the Mach number  at  stalling height, as a function of $\mathcal{L}$, the mechanical luminosity (which scales as $N_{OB}$).  Interestingly, superbubbles with Mach number (at stalling height) of order less than unity can be triggered by even a single SN. These, in principle, can accelerate later and therefore breakout of the disk. However, as we shall see later with our simulations, there is a minimum number of SNe needed for superbubbles to breakout of the disk, particularly for high density disks. We  also find  from Fig \ref{fig:komp_res} that in order
to achieve a Mach number at   stalling height of order $\sim 5$,  one needs $\mathcal{L} \ge 7 \times 10^{38}$ erg
s$^{-1}$, for $n_0=1$ cm$^{-3}$ and $z_0=500$ pc. This is larger than the estimate of Koo \& McKee (1992), and Mac Low \& McCray (1988), because of the inclusion of radiative loss from the shell. 
 If we consider $v_{z,{\rm min}}/c_s \ge 5$ as the breakout condition, then we find that larger densities and scale heights put more stringent condition on the bubble to breakout.

\begin{figure}
\centerline{
\includegraphics[scale=0.35,angle=-90]{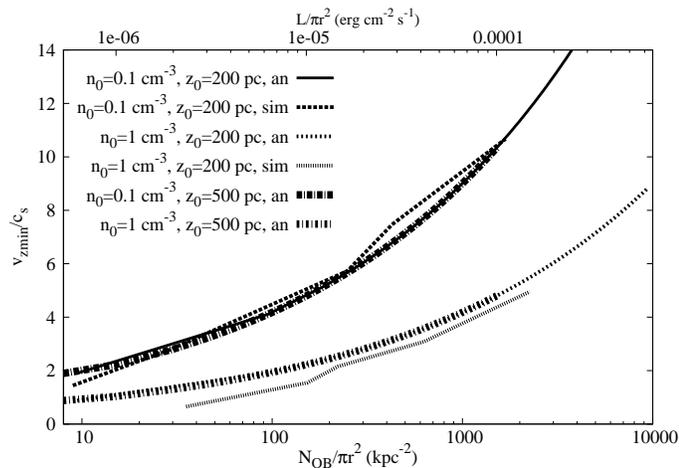}
}
{\vskip-3mm}
\caption{
Mach number of the top of the bubble  at the minimum velocity point is plotted as a function of $N_{OB}$ divided by the cross-sectional area of the bubble at the  stalling height, for analytical results and for Kompaneets  simulations. Analytical results are shown for
different values of mid-plane gas number densities $(1,0.1)$ cm$^{-3}$ and scale heights ($200,500$) pc, whereas simulation results for Kompaneets runs are shown for $n_0=0.1,1$
cm$^{-3}$ and scale height $z_0=200$ pc.
}
\label{fig:komp_sim_res}
\end{figure}

Next we plot in Figure \ref{fig:komp_sim_res} the minimum Mach number as a function of the surface density of $N_{OB}$,
considering the surface area of the bubble at the  stalling height.  Note that we are not concerned with the mean surface density of SFR in the disk galaxy here. The energy injection considered here is localized, but the relevant surface area as far as an emerging superbubble is concerned, is the area of the bubble in the plane of the disk at the point of breaking out. 
We find that for the surface density of energy deposition the  analytic curves become independent of the  scale height and depend only on the gas density and number of SNe. This is because the area of a superbubble in the plane parallel to the disk, scales with $z_0^2$, and is a constant for a given scale height.  We find that for a scale height of $500$ pc, the threshold surface density of SNe is $N_{OB} \sim 1000$ kpc$^{-2}$.

\section{Numerical Simulations}
In addition to analytic estimates and approximate calculations, we have performed 2-D axisymmetric hydrodynamic simulations of breakout using the ZEUS-MP code (Hayes et al. 2006). ZEUS-MP is a publicly available, second-order accurate Eulerian hydrodynamics code. We have carried out two sets of simulations: the first set compares numerical simulations with the analytic Kompaneets calculation of strong shocks in stratified atmospheres (hereafter these runs are referred to as `Kompaneets
runs'); and the second set of calculations use a more realistic setup, such as disk gravity, mass-loading of the ejecta, for shock (superbubble) breakbout in starforming galaxies (hereafter these runs will be called `realistic runs').

In this section we introduce the equations that we solve numerically, the initial and boundary conditions, and the choice of setup parameters. The simulations are run using the 2-D axisymmetric, spherical polar $(r, \theta, \phi)$ coordinates.

\subsection{Governing Equations}
We solve the following standard Euler's hydrodynamic equations including cooling, external gravity, and mass and energy loading at inner radii.
\begin{equation}
{d\rho \over dt}=-{\rho \nabla . \mathbf{v}}\, + S_\rho (r) \,,
\label{eq:simul1} 
\end{equation}
\begin{equation}
 \rho {d \mathbf{v} \over dt}=- \nabla p +\rho \mathbf{g} \, ,
 \label{eq:simul2}
\end{equation}
\begin{equation}
 {de \over dt}= -q^{-} (n,T) + S_e (r) \,,
 \label{eq:simul3}
\end{equation}
where $d/dt \equiv \partial / \partial t+ \mathbf{v}. \nabla$ is the Lagrangian derivative, 
$ \rho$ is the mass density, $\mathbf{v}$ is the fluid velocity, $p$ is the thermal pressure, $e=p/( \gamma -1)$  is the internal energy density (we use $ \gamma=5/3$ valid for an ideal non-relativistic gas), 
$\mathbf{g}=-{\rm sgn}(z) g \mathbf{\hat{z}}$ (sgn[$z$]=$\pm1$ for $z\gtrless0$) is the  constant external gravity  pointing towards the $z=0$ plane, $q^-\equiv n_en_i \Lambda(T)$ is the cooling term due to 
radiation where $n_e$ and $n_i$ are the electron and ion number densities, $\Lambda (T)$ is the cooling function (as given in eq \ref{eq:cooling_fn_formula}). There are source terms in the mass and internal energy equations ($S_\rho$, $S_e$). These terms are non-zero and constant only within 
$r_{\rm in}$, a small injection radius within which supernovae pump mass and energy into the interstellar medium. Note that the mass loading ($S_\rho$) and external gravity ($\mathbf{g}$) terms are used only for the realistic simulations and are set to zero for the Kompaneets runs.

The energy source function is chosen to mimic the energy input by supernovae, $S_e=\mathcal{L}/[(4 \pi/3) r_{\rm in}^3]$, where $\mathcal{L}=E_{SN} N_{OB}/t_\ast=6.3\times 10^{35}N_{OB}$ erg s$^{-1}$ is the 
supernova heating rate, $E_{SN}=10^{51}$ erg, $t_\ast=50$ Myr is the average lifetime of main sequence OB stars, and $N_{OB}$ is the number of OB stars. The mass-loading source function $S_\rho$ is chosen
as $S_\rho = \dot{M}/[(4 \pi/3) r_{\rm in}^3]$ where  $\dot M=\beta R_f ({\mathcal{L} /4\times 10^{41} {\rm erg~s}^{-1}}){M_\odot {\rm yr}^{-1}}$; where $R_f$ is the return-fraction ($=0.3$) and $\beta=3$,  that includes the effect of stellar winds, as inferred by Strickland \& Heckman (2009) in the case of M82. 
Tables \ref{table:table1} and \ref{table:table2} list the parameters for our Kompaneets and realistic simulations respectively.

 We implement energy injection by assuming the deposited energy to be thermalized within a radius $r_{\rm in}$, which we determine from the condition that the corresponding analytic solution of superbubble radius
 enters the Sedov-Taylor phase.  We assume a mass loading  $(1/2) \, N_{OB} \, 10 $ M$_{\odot}\sim 5 \, N_{OB}\, $ M$_{\odot}$, for a typical mass of OB stars of order $10$ M$_{\odot}$ and half the 
 progenitor mass being ejected during the supernova. This is an approximation, however we found that mass loading at this level has negligible effects on the evolution of superbubbles.
 The superbubble enters the Sedov-Taylor phase when the ejecta mass equals the mass swept up by the shell. We choose this radius to be our $r_{\rm in}$ because before this phase, most of the energy of the superbubble is in kinetic form, and the assumption of most of the energy being thermalized is appropriate only in the Sedov-Taylor phase. Moreover, $r_{\rm in}$ should be smaller than the radius at which the shock becomes radiative.  Tables \ref{table:table1} and \ref{table:table2} show the values of $r_{\rm in}$ for different simulations.

\subsection{Initial and Boundary conditions}
We have used ZEUS-MP in spherical polar $(r,\theta,\phi)$ coordinates. We fix the inner radial boundary (the mass and energy injection radius) at $r_{\rm min} < r_{\rm in}$, and the outer boundary
at $r_{\rm max}=3\hbox{--}10$ kpc, depending on the distance reached by the superbubble in $1.23 \times 10^{15}$ s ($39.3$ Myr) , the maximum time for which we run the simulations. For $\theta-\phi$ coordinates, $\theta$ goes from $0$ to $\pi$, and $\phi$ goes from $0$ to $2 \pi$. We use a logarithmically spaced grid in the radial direction such that there are equal number of grid points in $[r_{\rm min},(r_{\rm min}r_{\rm max})^{1/2}]$ and $[(r_{\rm min}r_{\rm max})^{1/2}, r_{\rm max}]$; the grid is uniformly spaced in the other directions. 
The resolution adopted for our simulation is $512 \times 256 \times 1$, in the $r, \theta, \phi$ directions (although we have used a higher resolution of $1024 \times 1024 \times 1$ for our study
of thermal instability in the relevant cases).
Outflow boundary conditions are applied at the outer radial boundary.  Inflow-outflow boundary condition is applied at the inner radial boundary such that mass is allowed to leave or enter the box. Reflective boundary conditions are imposed at $\theta=0,~\pi$, and periodic boundary conditions are applied in the $\phi$ direction.

The initial conditions in Kompaneets and realistic runs are different. Both set of runs have an initial temperature of $10^4$ K corresponding to the stable WIM. In Kompaneets runs the density is stratified in the vertical ($z-$) direction as $ \rho(t=0) \propto e^{-z/z_0}$, where $z_0$ is the scale height. Thus, the initial state is not in dynamical equilibrium. However, since the sound speed is very small, the evolution occurs because of fast energy injection in the center. We have verified that the results are the similar as for simulations with a constant initial pressure.  For the Kompaneets runs, we have used a scale height of $z_0=200$ pc.

For realistic runs the initial ISM is symmetric with respect to the vertical direction, with $ \rho(t=0) \propto e^{-|z|/z_0}$, where the scale height is determined self-consistently for an isothermal gas in hydrostatic equilibrium; i.e., the strength of the constant gravitational acceleration is chosen to be $g = c_s^2/ z_0$ where $c_s \equiv \sqrt{kT/\mu m_p}$ ($k$ is Boltzmann constant, $\mu$ is the mean particle mass, and $m_p$ is proton mass) is the isothermal sound speed and $z_0$ is the scale height.

\begin{table}
\caption{Parameters for Kompaneets runs ($\mathcal{L}=6.3 \times 10^{35}$ erg s$^{-1} \,N_{OB}$)}
\centering
\begin{tabular}{c c c c c}\\[0.5ex]
\hline\hline
$n_0$ (cm$^{-3}$) & $N_{OB}$ & $r_{\rm in}$ (pc) & $r_{\rm min}$ (pc) & $r_{\rm max}$ (pc)\\ [0.5ex]
\hline
 0.1 & 1   & 10 & 5 & 2500\\
 0.1 & 10  & 21 & 10 & 2500\\
 0.1 & 100  & 44 & 30 & 2500\\
 0.1 & 300   & 63 & 40 & 3000\\
 0.1 & 1000   & 94 & 70 & 3500\\
 1.0 & 1  & 5 & 3 & 2500\\
 1.0 & 10  & 10 & 5 & 2500\\
 1.0 & 100  & 20 & 10 & 2500\\
 1.0 & 300   & 29 & 15 & 3000\\
 1.0 & 1000  & 44 & 30 & 3000\\[1ex]
\hline
\end{tabular}
\label{table:table1}
\end{table}

\subsection{Kompaneets runs}
We first describe the results of our Kompaneets runs, of superbubbles in a stratified atmosphere without external gravity or mass-loading. Figure \ref{fig:komp_sim_res} shows the variation of the minimum Mach number of the top of the superbubble as a function of the surface density of energy injection
in the disk, for a scale height of $200$ pc and two values of ambient density,
$n_0=0.1$ and $1$ cm$^{-3}$. We find that the analytical results overestimate the Mach number of the superbubbles 
compared to the simulations by a factor of order $\sim 1$ for the case of large ambient density ($1$ cm$^{-3}$), because 
the analytical estimate of energy loss described in the previous section is based on simplified assumptions.  Note that since we
determine the value of $z_+$ by the position of the maximum density, clumps in the shell formed due to thermal instability (see below for details) introduce
some uncertainity. This manifests in the kinks seen in the simulation results in Fig. 4 and also later in Fig. 6.

\subsection{Realistic runs}
Next we describe  simulations that includes vertical disk gravity and mass loading.  We study the case of ambient gas 
at $T=10^4$ K, with mid-plane densities $n_0=0.1$ and $1$ cm$^{-3}$, and scale heights $z_0=100$
and $500$ pc. 

Our choice of parameters essentially brackets the possible range of gas density and scale height in disk galaxies. For example,
the distribution of the extraplanar gas in Milky Way has two components, that of warm 
ionized gas and cold HI. The warm ionized gas has been observed to have an exponential profile with $n_0 \sim 0.01\hbox{--}0.03$ cm$^{-3}$
and $z_0\sim 400\hbox{--}1000$ pc (Reynolds 1991; Nordgren \etal 1992; Gaensler \etal 2008). 
For HI distribution, Dickey \& Lockman (1990) found that the vertical
distribution 
is best described by a Gaussian with  
FWHM of $230$ pc and a central density of $0.57$ cm$^{-3}$. The combined distribution of these two components are bracketed
by exponentials with the scale heights and mid-plane densities assumed here. 

\begin{table}
\caption{Parameters for Realistic runs}
\centering
\begin{tabular}{cccccc}\\
\hline\hline
$z_0$ (pc) & $n_0$ (cm$^{-3}$) & $N_{OB}$ & $r_{\rm in}$ (pc) & $r_{\rm min}$ (pc) & $r_{\rm max}$ (pc)\\ 
\hline
 100 & 0.1 & 1   & 10 & 5 & 1000\\
 100 & 0.1 & 10  & 21 & 10 & 2500\\
 100 & 0.1 & 100  & 44 & 10 & 2500\\
 100 & 0.1 & 300   & 63 & 10 & 2500\\
 100 & 0.1 & 1000   & 94 & 50 & 2500\\
 100 & 1 & 100  & 20 & 10 & 2500\\
 100 & 1 & 300   & 29 & 15 & 2500\\
 100 & 1 & 1000  & 44 & 30 & 2500\\
 100 & 1 & 2000   & 55 & 40 & 2500\\
 100 & 1 & 3000  & 63 & 40 & 2500\\
 500 & 0.1 & 10  & 21 & 10 & 2500\\
 500 & 0.1 & 100  & 44 & 10 & 2500\\
 500 & 0.1 & 300   & 63 & 30 & 3500\\
 500 & 0.1 & 1000   & 94 & 50 & 3500\\
 500 & 0.1 & 3000   & 135 & 110 & 3500\\
 500 & 0.1 & 10000   & 201 & 160 & 3500\\
 500 & 0.1 & 50000   & 344 & 300 & 12000\\
 500 & 0.1 & 100000   & 433 & 400 & 12000\\
 500 & 1 & 1000  & 44 & 30 & 2500\\
 500 & 1 & 2000   & 55 & 40 & 2500\\
 500 & 1 & 3000  & 63 & 40 & 2500\\
 500 & 1 & 5000  & 75 & 50 & 3500\\
 500 & 1 & 10000  & 94 & 70 & 3500\\
 500 & 1 & 100000  & 201 & 150 & 5500\\
\hline
\end{tabular}
\label{table:table2}
\end{table}

We also use smaller scale heights in our simulations. The scale height near the centres of galaxies is smaller than that in the outer regions, because of deeper gravitational potentials in the central regions. Also Dalcanton, Yoachim, Bernstein (2004) found that the HI scale height of disk galaxies varies with the rotation speed (or, equivalently, the galactic mass). Dwarf spirals with rotation speed $\sim 50$ km s$^{-1}$ have $z_0\sim 200$ pc whereas larger galaxies (with rotation speed in excess of $120$ km s$^{-1}$) have $z_0=500\hbox{--}1000$ pc.  Also, as Basu \etal (1999) have found, the scale height encountered by Milky Way superbubbles such as W4 is rather small
 ($\le 100$ pc).

We first find that unlike in the analytical case, where superbubbles ultimately break out of the disk sooner or later, irrespective of the energetics, the realistic simulation runs show that for high density disk 
material ($n_0\ge 1$ cm$^{-3}$) , superbubbles keep decelerating for ever for  a surface density of OB stars $\sim 100 (z_0/100 \, {\rm pc})$ kpc$^{-2}$. In other words superbubbles never break out of the disk in these cases. The corresponding energy injection surface density is $\sim 2\hbox{--}5 \times 10^{-5}$ erg cm$^{-2}$ s$^{-1}$ For lower density ambient gas, $n_0\sim 0.1$ cm$^{-3}$, however, even a single SN event can drive a bubble through the disk. We note that this limit is consistent with that found by Silich \& Tenorio-Tagle (2001) for a Milky Way type disk.

\begin{figure}
\centerline{
\epsfxsize=0.5\textwidth
\epsfbox{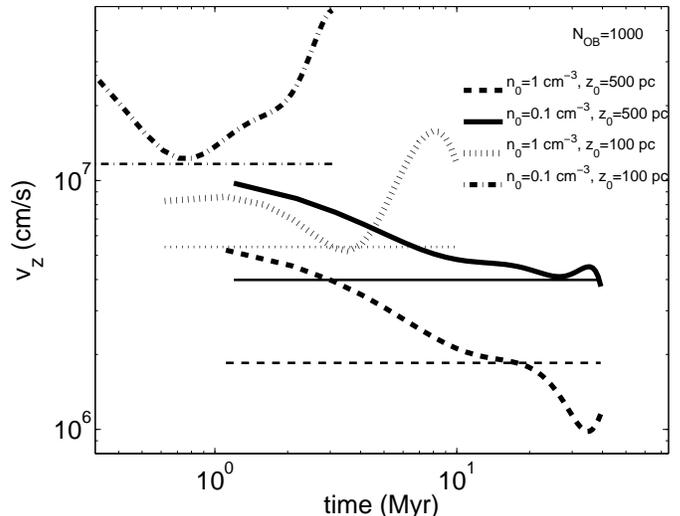}
}
{\vskip-3mm}
\caption{
Velocity of the topmost point of the bubble is plotted against time for $N_{OB}=1000$, but for different combinations of scale height ($z_0=100,500$ pc) and mid-plane gas density ($n_0=0.1,1$ cm$^{-3}$). The horizontal lines in each case shows $(1/5) (\mathcal{L} /\rho _0 z_0^2)^{1/3}$, the expected scaling.
}
\label{fig:vzminvst}
\end{figure}

 In the case of a superbubble breaking out of the disk, there are differences in the way they evolve depending on the energy injection rate.
 We show the evolution of the speed of the topmost point of the bubble as a function of time
for four cases in Figure \ref{fig:vzminvst}, for two mid-plane densities ($n_0=0.1, 1$ cm$^{-3}$),
and two scale heights ($z_0=100,500$ pc), all for  a surface density of OB stars of $1000$ kpc$^{-2}$. The curves show that the bubbles show acceleration after breakout of the disk only for the case of low density and small scale height ( see the curve at the top-left corner, for $n_0=0.1$ cm$^{-3}$, $z_0=100$ pc). In other cases,  for disk column density $\ge 3 \times 10^{19}$ cm$^{-2}$, the bubbles either coast along
with the the speed that they reach at the breakout, or decelerate to some extent, for a considerable period of time before they start accelerating after reaching a distance of several scale heights. The curves show that the speed at  the stalling height, or the minimum speed of the bubbles, is an important characteristics of the bubble dynamics. It is important because this is the characteristic speed with which the bubble sweeps most of the extra-planar region of the halo. Also, since the bubble begins to accelerate only after reaching a distance of a few times the scale height, the corresponding
Rayleigh-Taylor instability should not set in at the scale height, but at a much larger distance. We
shall re-visit this point in the next section on instabilities.  In some cases, the curves show a deceleration at late times. This is due to the formation of clumps in the shell from radiative cooling, which often sink through the hot gas owing to gravity.

We have found that typically the minimum speed $v_{z,{\rm min}} \sim (1/5) ( \mathcal{L} /
\rho_0 z_0^2) ^{1/3} \sim z_0/(5 t_d)$, where $t_d$ is the dynamical time defined earlier. These values are shown as horizontal lines in Figure \ref{fig:vzminvst} for respective cases. It is easy to see that in case of little radiation loss, the speed of the bubble at the time of reaching the scale height is 
$\sim (3/5) (\mathcal{L}/\rho_0 z_0^2)^{1/3}$, as expected from the self-similar evolution of a bubble ($r\sim (\mathcal{L} t^3/ \rho_0)^{1/5}$). Our simulations show that the actual speed is roughly a third of this 
value, and therefore shows the importance of radiative loss in the dynamics of superbubbles. As analytically derived earlier, radiation losses remove as much as 70\% of the total energy of the superbubbles. 
We recall that for an ambient medium with a given temperature, the dimensionless quantity defined by Mac Low \& Norman (1988) is  $D \sim (5 v_{z,{\rm min}}/c_s)^3$, so that their condition of $D\ge 100$ for break out corresponds to a minimum Mach number of order unity.

\begin{figure}
\centerline{
   \includegraphics[scale=0.35,angle=-90]{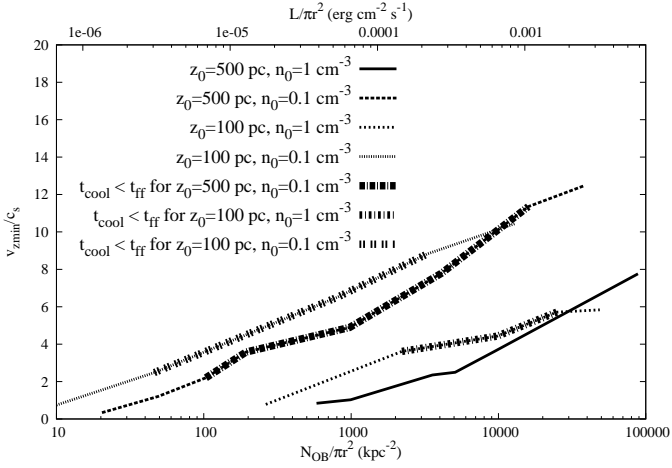}
}
{\vskip-3mm}
\caption{
The minimum Mach number of the top of the bubble in our realistic runs are shown as a function
of $N_{OB}$ per kpc$^{-2}$, and $\mathcal{L}/\pi r^2$ (erg cm$^{-2}$ s$^{-1}$), for $n_o=0.1,1$ cm$^{-3}$ and $z_0=100, 500$ pc.  Note that, for $n_0=1$ cm$^{-3}$, the shocks stall for a surface
density of OB stars $\le 500$ kpc$^{-2}$. The cases for which $t_{cool} < t_{ff}$, are shown by darkened points, these cases are marked by thermal instability.
}
\label{fig:real_sim1}
\end{figure}

We show the resulting value of minimum Mach number of superbubbles for different $n_0$ and
$z_0$ in Figure \ref{fig:real_sim1}, as a function of  surface density of energy injection. The curves show that in terms of energy injection or SNe surface density, the
crucial parameter is the mid-plane gas density, which separates the curves, as was also indicated by our analytical results. Superbubbles with a given surface density of energy injection find it easier to break out of disks with lower mid-plane density. However, scale height also makes a  small difference unlike in the analytical calculations; a higher energy density is required to clear a thicker disk.

The important features of our results as shown in Figure  \ref{fig:real_sim1} are:
\begin{itemize}
\item As mentioned above, the condition for a break out from a dense ambient medium with gas density of $n_0=1$ cm$^{-3}$ is an energy injection rate surface density of $2\hbox{--}5 \times 10^{-5} $ erg cm$^{-2}$ s$^{-1}$. For lower gas densities, the required rate density is $\sim 10^{-6}$ erg cm$^{-2}$ s$^{-1}$. The corresponding Mach number for these superbubbles that can be as low as of order unity.
\item Superbubbles that can break out with a larger Mach number of $\sim 3\hbox{--5}$ corresponds to $\sim 1000$ $N_{OB}$
kpc$^{-2}$, or an energy injection surface density of $10^{-4}$ erg~s$^{-1}$ cm$^{-2}$, for the most realistic  spiral disks, with $n_0=0.1$ cm$^{-3}$, $z_0=500$ pc, or $n_0=1$ cm$^{-3}$, $z_0=100$ pc (which for Milky Way case describes either the warm extra-planar or cold gas). We note that the largest OB associations have $\sim 10^4$ M$_{\odot}$ (McKee \& Williams 1997), and with a cross-sectional area of order $\pi \, (\pi z_0)^2$ (since $r_{\rm max}\sim \pi z_0$ asymptotically; see
equation \ref{eq:rmax}), a superbubble blown by such a large OB association can  only have $\le 10^3$ SNe kpc$^{-2}$, for a Salpeter IMF. Therefore we can conclude that (a) only the largest of the OB associations can produce a bubble that can break out of Milky Way-type disks, (b) this 
also corresponds to the minimum energy injection rate of $10^{-4}$ erg~s$^{-1}$ cm$^{-2}$ as
observed by Dahlem \etal (1995)  for the existence of radio emitting halo gas, and (c) larger ISM density or scale height would require more than one OB association to produce a 
superbubble or adjacent multiple bubbles that can coalesce and grow together.  Recent simulations show that cosmic rays can stream through ISM gas to considerable heights above the
disk, and break out of superbubbles can provide such channels (Uhlig \etal (2012).
\item As explained earlier, the minimum speed of $3\hbox{--}5 \, c_s\sim 30$ km s$^{-1}$, for 
an ambient gas at $10^4$ K, also corresponds to the case where the hot (and multiphase gas;
see next section on instabilities and gas cooling) interior gas can sweep up to a height of $\sim 1$
kpc within a time period of $\sim 50$ Myr, the time scale over which OB stars explode and keep injecting energy in the bubble. Combined with the result mentioned above, we can conclude that 
an energy injection rate of $10^{-4}$ erg~s$^{-1}$ cm$^{-2}$, or $\sim 1000$ $N_{OB}$ kpc$^{-2}$ can not only produce a bubble that can break out of the disk but also fill the halo up to a height 
of order $\sim 1$ kpc.
\item If we insist on a larger Mach number at stalling height, to be $5\hbox{--} 10$, then the
energy injection rate becomes $\sim 10^{-3}$ erg~s$^{-1}$ cm$^{-2}$, with $\sim 2\times  10^4$ $N_{OB}$ kpc$^{-2}$. Using a time scale of $\sim 50$ Myr   of OB stars,   the corresponding SFR surface density for  a Salpeter IMF is $\sim 0.06$ M$_{\odot}$ yr$^{-1}$ kpc$^{-2}$.   If superbubbles seed galactic outflows, then the gas speed is required to be a few hundred km s$^{-1}$, and the Mach number at stalling height is needed to be much larger than ten, and the corresponding requirement on SFR surface density increasing to $\sim 0.1$ M$_{\odot}$ yr$^{-1}$ kpc$^{-2}$, the observed threshold.
Therefore, the Heckman (2000)  threshold
( $\sim 0.1$ M$_{\odot}$ yr$^{-1}$ kpc$^{-2}$) for superwinds corresponds to a larger requirement on the part of superbubbles, of not only 
breaking out of disks but doing so with a large Mach number.
\end{itemize}

\section{Thermal and Rayleigh-Taylor instability}
The focus till now was on the important $v_z/c_s$ parameter (the minimum Mach number of the shell) which determines the fate of the superbubble after it crosses the scale height. In this section we discuss the role of different instabilities, in particular Rayleigh-Taylor and thermal instabilities, in our 2-D breakout simulations. When the superbubble reaches about a scale height, the shock is generally believed to accelerate owing to the decrease in pressure. This should lead to the onset of the Rayleigh-Taylor (RT) instability, as has been invoked in previous analytical works (e.g., Koo \& McKee 1992) and seen in numerical simulations (e.g., Mac Low, McCray \& Norman 1989). 
However, as mentioned earlier, our simulations show that superbubbles {\it do not} accelerate until after they reach a distance of several scale heights (as was also suggested by Ferrara \& Tolstoy 2000 
who assumed spherical bubbles). Therefore RT instability occurs at a distance {\it much larger} than the scale height.  Also we find that before the onset of RT 
instability, the superbubble expanding in the disk suffers from thermal instability in the early stages of its evolution. This instability leads to clumping and fragmentation of the shell of the superbubble well in advance of the  RT instability, and can therefore affect the outcome of the RT instability.

\begin{figure}
\centerline{
\epsfxsize=0.5\textwidth
\epsfbox{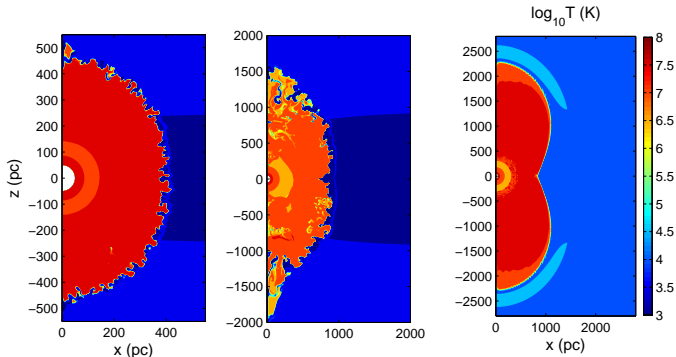}
}
{\vskip-3mm}
\caption{
Temperature contours (colour coded) for a superbubble with $N_{OB}=5000$,
$n_0=1$ cm$^{-3}$, $z_0=500$ pc, at $t=9$ Myr, when the top of the bubble has reached a distance of the scale height (left panel),  at $39.3$ Myr, when it has reached a distance $\sim 3 z_0$ (middle panel). The rightmost panel shows the case of the same superbubble without radiative cooling at $t=39.3$ Myr, the same evolutionary epoch as the middle panel.
}
\label{fig:sim_temp}
\end{figure}

\begin{figure}
\centerline{
\epsfxsize=0.5\textwidth
\epsfbox{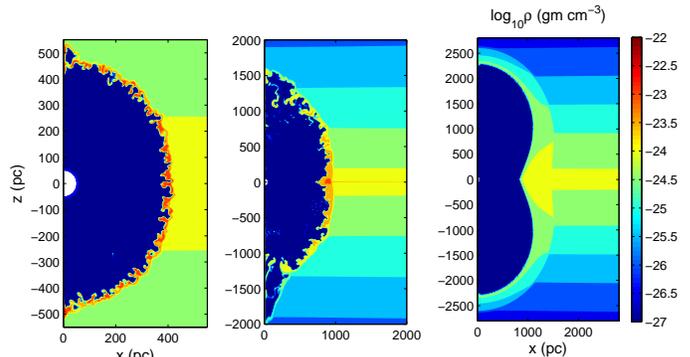}
}
{\vskip-3mm}
\caption{
Density contours for the same cases as in Fig \ref{fig:sim_temp}. Here, fragmentation of the shell is clearly seen in the run with cooling.
}
\label{fig:sim_den}
\end{figure}

Figure \ref{fig:sim_temp} shows the 2-D snapshots of temperature at two different times for our fiducial high resolution run ($N_{OB}=5000$, $n_0=1$ cm$^{-3}$, $z_0=500$ pc). Figure \ref{fig:real_sim1} indicates that the minimum Mach number for this case is $\approx 2$ and the bubble is just about able to break out within the starburst timescale. The temperature snapshot at early time (9 Myr), when the bubble has  just reached the scale-height, shows that the bubble is roughly spherical. The radiative shell seems to develop corrugations where the hot bubble gas and the radiatively cooled shocked gas interpenetrate. The shell is at $\approx 10^4$ K (the same as the ambient ISM temperature), the temperature below which the cooling function drops suddenly and the gas becomes thermally stable. The dense shell is more clearly seen in the density snapshots of Figure \ref{fig:sim_den}. The corrugations are definitely driven by radiative cooling because the run without radiative cooling shows a smooth shell (the 
third panel in Figs. \ref{fig:sim_temp} and \ref{fig:sim_den}).

While the fragments of cold shell are confined to the bubble boundary at early times, the cold gas lags behind the hot gas at later times because the hot gas is pushed out  by supernova heating. The cold blobs are only pushed out because of the drag force due to the hot gas but eventually trail behind. The cold blobs embedded in the hot gas are reminiscent of the cold multiphase filaments observed in  galactic outflows, such as M82. Since in our simulations cold gas leaves the simulation box from the inner boundary, all the cold blobs embedded in the hot bubble come from the fragmenting cold shell. In reality, some cold gas from the cold star-forming regions can also be uplifted by the hot gas. At late times, in the runs with cooling, there are some signs of bubble breaking out because of RT instability close to the polar regions. All such signatures of RT instability are missing in the run without cooling (panel 3). This is mainly because RT instability in the run with cooling is seeded with large amplitude 
perturbations by corrugations caused by shell cooling.

\begin{figure}
\centerline{
\epsfxsize=0.55\textwidth
\epsfbox{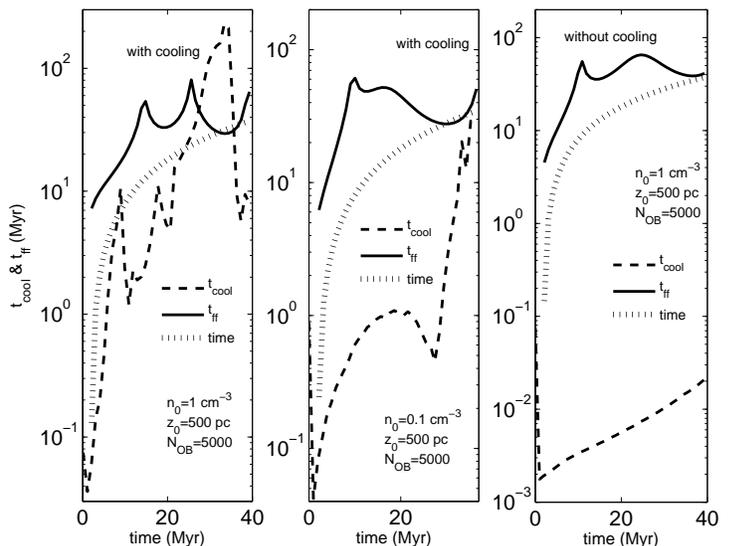}
}
{\vskip-3mm}
\caption{
The free-fall and cooling timescales for the shell material are plotted against time, for two examples with $N_{0B}=5000$, $z_0=500$ pc, and $n_0=1$ cm$^{-3}$ (left panel), $n_0=0.1$ cm$^{-3}$
(middle panel). The grey lines show the time elapsed in each cases for comparison. The right panel shows the case of no radiation cooling for $n_0=1$ cm$^{-3}$. The leftmost and rightmost panels correspond to the runs shown in Figs \ref{fig:sim_temp} and \ref{fig:sim_den}.
}
\label{fig:tff-tcool}
\end{figure}

In order to assess the relative importance of thermal and RT instabilities, we compare the two times scales in 
Figure \ref{fig:tff-tcool}. We note that the time scale for 
RT instability ($t_{RT} =\sqrt{1/(\dot{v_z}+g) k}$) is comparable to the free-fall time  
($t_{\rm ff}=\sqrt{2 z/(\dot{v_z}+g)}$, for the largest mode with $k \sim 2\pi/z$, where $z$ and $\dot{v_z}$ are the height and acceleration of the shell, and 
$g$ is the acceleration due to gravity. We plot this time scale with a solid line in Figure \ref{fig:tff-tcool}, along with 
the cooling time ($t_{\rm cool}=1.5 kT/n \Lambda$) of the shell as a function of time for runs corresponding to Figures \ref{fig:sim_temp} \& \ref{fig:sim_den}.  We use the position 
of the outermost densest part to identify the shell position. In the left panel of Figure \ref{fig:tff-tcool}, we show the case of $N_{OB}=5000$, $n_0=1$ cm$^{-3}$, $z_0=500$ pc. We expect the shell to cool radiatively if $t_{\rm cool}$ is shorter than time. And indeed, the radiative cooling time is shorter than time at early times. This is consistent with the cooling and fragmentation of the dense shell seen in Figs. \ref{fig:sim_temp} \& \ref{fig:sim_den}. One point of caution: we should ideally plot the cooling time of the shell assuming  the shell temperature and density corresponding to an adiabatic shock because cooling will happen if this timescale is short. Here we are plotting the cooling time of the shell, which for the left panel case, has already cooled to low temperatures. Since cooling time increases sharply 
below $10^4$ K, $t_{\rm cool}$ is barely smaller than time in the left panel of Figure \ref{fig:tff-tcool}. At later times $t_{\rm cool}$ becomes longer than time and we do not expect the newly accumulated shell material to cool.  The RT timescale ($\approx t_{\rm ff}$) is always longer than time for the fiducial run.  The free-fall time increases initially as the shock slows down until a scale height. After that the shock moves at a small Mach number $\sim 2$. This is consistent with the fact that we do not see vigorous RT instability in Figures \ref{fig:sim_temp} \& \ref{fig:sim_den}.

The  middle panel of the Figure \ref{fig:tff-tcool} shows various timescales for a midplane density of $n_0=0.1$ cm$^{-3}$. The cooling time for this case is shorter than the cooling time for the higher density case. This seems inconceivable given the higher density and efficient cooling for the run in the left panel. This discrepancy arises because although the density for the $n_0=0.1$ cm$^{-3}$ is smaller, the temperature of the post-shock gas is $10^5$ K, where the cooling function peaks. Consequently the cooling time is shorter than the higher density run.  For comparison, we
have also plotted the cooling and free-fall timescales for the runs without cooling in the right
panel. The density and temperature snapshots for this run do not show cooling-induced fragmentation.

We have shown in Figure \ref{fig:real_sim1} by darkened points the cases in which $t_{\rm cool}$ is
always less than $t_{\rm ff}$, for different values of $\mathcal{L}/\pi r^2$, $z_0$ and $n_0$. We find that these cases mostly appear for which, roughly,  $10 \ge v_{z,{\rm min}} \ge 3$, except for the case of $z_0=500$ pc and 
$n_0=1$ cm$^{-3}$, for which there is a cross-over point in time after $t_{\rm cool}\ge t_{\rm ff}$. We note that this range of $v_{z,{\rm min}}$ corresponds to a case in which the shell temperature ($T_s$)  remains
in the range of $2 \times 10^4 \le T_s \le 10^6$, where the cooling function peaks. This implies a range in $N_{OB}$ for which thermal instability is imporant. In the low $N_{OB}$ limit, the shock is not strong enough and $T_s\le 10^4$ K, and in the high $N_{OB}$ case, the shock is very strong
($T_s >10^6$ K) and $t_{\rm ff}$ (RT timescale) is shorter than $t_{\rm cool}$ at late times.

We are therefore led to conclude that superbubbles are affected not only by RT instability but also by thermal instability, depending on the density and energy injection. This implies that the fragmentation of the bubble shell that releases the hot interior gas into the halo occurs under the combined effects of thermal instability at early times and RT instability at late times if the Mach number at stalling epoch is large enough.

\section{Discussion \& Summary }
Superbubbles with fragmented shells are believed to ultimately form `chimneys' (Norman \& Ikeuchi  (1989), which connect the halo gas to the processes in the disk in different ways. Apart from transporting hot gas to the halo, chimneys provide a natural channel for Lyman continuum photons from hot stars in the disk to reach the diffuse ionized medium of the Reynolds layer (Reynolds 1991; Dove \& Shull 1994). It is however important for the superbubble shells to fragment before the main sequence life times of O stars for a substantial fraction of ionizing radiation to escape the disk (Dove, Shull, Ferrara 2000). This implies a fragmentation time scale of $\sim 3\hbox{--}5$ Myr, which is comparable to the dynamical timescale ($t_d \sim z_0^{5/3} (\rho_0 / \mathcal{L})^{1/3}$), for superbubbles with $\mathcal{L} \sim 10^{38}$ erg (corresponding to $N_{OB} \sim 200$), typical disk parameters. This is the energy scale for the largest of the OB associations, and as our results show superbubbles with smaller 
energetics find it hard to pierce through the disk, unless the OB association is located much above the mid-plane level.

In other words, for superbubbles to act as effective conduits of ionizing radiation for the halo, or for the intergalactic medium (at high redshift, in the context of the epoch of reionization), the superbubbles need to fragment roughly around the time when they reach a scale height. This is unlikely to happen only through RT instability as superbubbles do not accelerate until reaching a distance of several scale heights. Also, as de Avillez \& Breitschwerdt (2005) have discussed on the basis of simulations of a magnetized ISM, superbubble shells can stabilize against RT instability in the presence of magnetic fields. 
In this regard, the clumping of the shell from thermal instability at an early phase of evolution of the superbubble can be important.

We have studied the evolution of superbubbles in stratified disks analytically and with simulations. Our results can be summarised as follows:
\begin{itemize}
\item  Our analytic calculations show that radiation losses are important for superbubble dynamics. Radiation loss is more important for superbubbles with continuous energy injection than a supernova remnant of similar total energy.  
We estimate almost 70\% of the total energy being 
radiated away.  We have further checked our analytical results
with numerical simluations. We found that analytic results 
match the simulations well, differing at most by a factor of order unity for the case of large ambient density. The results obtained by the analytical means therefore provide a useful benchmark to compare with realistic simualtions.
 Also, for disks with large gas density, with $n_0\ge 1$ cm$^{-3}$, superbubble breakouts are not possible for surface density of OB stars $\le 100 (z_0/100 \, {\rm pc})$ kpc$^{-2}$, or an equivalent energy injection surface density of $\le (2 \hbox{--}5) \times 10^{-5}$ erg cm$^{-2}$ s$^{-1}$.
\item Superbubbles that  emerge from the disk with Mach number of order $2\hbox{--}3$ require 
an energy injection rate of $\sim 10^{-4}$ erg cm$^{-2}$ s$^{-1}$, corresponding to explosions triggered by the largest OB associations with $10^4$ M$_{\odot}$. This energy injection scale
corresponds to disk galaxies with synchrotron emitting gas in the extra-planar regions.
\item  Vigorous superbubbles that break out of the disk with sufficiently large Mach number  ($\ge 10$) , correspond to an energy injection rate of $\sim 10^{-3}$ erg cm$^{-2}$ s$^{-1}$, or
equivalently, a SFR surface density of $\sim 0.1$ M$_{\odot}$ yr$^{-1}$ kpc$^{-2}$. These
superbubbles require more than one OB associations to produce and sustain their dynamics, and
this energy injection scale corresponds to (a) the existence of multiphase gas in the halo of 
disk galaxies, and (b) the Heckman threshold for the onset of superwinds.
\item Superbubbles do not accelerate until reaching a vertical distance of a few scale heights
(of order $\sim 2$), which implies that RT instability helps to fragment the shells not at a
distance of a scale height but at a much larger height. Also, we find that for typical disk parameters, thermal instability acts on the shell at the early stages of superbubble evolution, and forms clumps and fragments in the shell, much before the shell is acted upon by RT instability.  Radiative cooling therefore manifests in seeding thermal instability, which has important implications for the clumping of superbubble shell and producing channels of leakage for ultraviolet radiation into the halo.
\end{itemize}

\bigskip
 We thank Sergiy Silich for helpful comments on a draft of the paper. We also thank an anonymous referee for the useful comments. This work is partly supported by an Indo-Russian project 
 (RFBR grant 08-02-91321, DST-India grant INT-RFBR-P121).

\end{document}